\documentclass[twocolumn,amsmath,amssymb,aps,secnumarabic,%
    nofootinbib,groupedaddress,floatfix]{revtex4}
\usepackage{times,mathptmx,amsthm,pstricks}
\newtheorem{theorem}{Theorem}[section]

\newtheorem{corollary}[theorem]{Corollary}
\newtheorem{proposition}[theorem]{Proposition}
\theoremstyle{definition}
\newtheorem{algorithm}{Algorithm}
\theoremstyle{remark}
\newtheorem*{remark}{Remark}

\newcommand{\C}{\mathbb{C}}
\newcommand{\Z}{\mathbb{Z}}
\renewcommand{\tensor}{\otimes}
\DeclareMathOperator{\Irrep}{Irrep}

\newcommand{\cE}{\mathcal{E}}
\newcommand{\cH}{\mathcal{H}}
\newcommand{\cM}{\mathcal{M}}
\newcommand{\ket}[1]{|#1\rangle}
\newcommand{\bra}[1]{\langle#1|}
\newcommand{\ketpsi}[1]{\ket{\psi_{#1}}}
\newcommand{\braket}[1]{\langle#1\rangle}
\newcommand{\st}{\bigm|}
\newcommand{\tO}{\widetilde{O}}
\newcommand{\ceil}[1]{\lceil #1 \rceil}
\newcommand{\bceil}[1]{\bigl\lceil #1 \bigr\rceil}

\newcommand{\injects}{\hookrightarrow}

\newcommand{\ie}{\textit{i.e.}}

\newcommand{\thm}[1]{Theorem~\ref{#1}}
\renewcommand{\sec}[1]{Section~\ref{#1}}

\newcommand{\cor}[1]{Corollary~\ref{#1}}
\newcommand{\prop}[1]{Proposition~\ref{#1}}
\newcommand{\alg}[1]{Algorithm~\ref{#1}}
\newcommand{\eq}[2]{\begin{equation}\label{#1}#2\end{equation}}

\newcommand{\fig}[1]{Figure~\ref{#1}}
\newcommand{\tab}[1]{Table~\ref{#1}}

\newenvironment{fullfigure}[2]
    {\begin{figure}[htb]\begin{center}\def\fullfiga{#1}\def\fullfigb{#2}}
    {\caption{\fullfigb.}\label{\fullfiga}
        \end{center}\end{figure}}
\newenvironment{fulltable}[3]
    {\begin{table}[htb]\def\fulltaba{#1}\def\fulltabb{#3}\begin{tabular}{#2}}
    {\end{tabular}\caption{\fulltabb.}\label{\fulltaba}\end{table}}

\psset{linewidth=.5pt,dash=4pt 4pt,arrowsize=2pt 3}
\SpecialCoor

\begin{document}
\title{A subexponential-time quantum algorithm for the dihedral hidden subgroup
    problem}

\author{Greg Kuperberg}
\email{greg@math.ucdavis.edu}
\thanks{Supported by NSF grant DMS \#0072342}
\affiliation{Department of Mathematics, University of
    California, Davis, CA 95616}

\begin{abstract}
We present a quantum algorithm for the dihedral hidden subgroup problem with
time and query complexity $2^{O(\sqrt{\log\ N})}$. In this problem an oracle
computes a function $f$ on the dihedral group $D_N$ which is invariant under a
hidden reflection in $D_N$. By contrast the classical query complexity of DHSP
is $O(\sqrt{N})$.  The algorithm also applies to the hidden shift problem
for an arbitrary finitely generated abelian group.

The algorithm begins as usual with a quantum character transform, which in the
case of $D_N$ is essentially the abelian quantum Fourier transform. This yields
the name of a group representation of $D_N$, which is not by itself useful, and
a state in the representation, which is a valuable but indecipherable qubit. 
The algorithm proceeds by repeatedly pairing two unfavorable qubits to make a
new qubit in a more favorable representation of $D_N$. Once the algorithm
obtains certain target representations, direct measurements reveal the hidden
subgroup.
\end{abstract}
\maketitle

\section{Introduction}
\label{s:intro}

The hidden subgroup problem (HSP) in quantum computation takes as input a group
$G$, a finite set $S$, and a black-box function (or oracle) $f:G \to S$. By
promise there is a subgroup $H \subseteq G$ such that $f(a) = f(b)$ if and only
if $a$ and $b$ are in the same (right) coset of $H$.  The problem is to
identify the subgroup $H$.  We assume that $G$ is given explicitly; black-box
groups are a separate topic \cite{IMS:instances}.

Shor's algorithm \cite{Shor:factorization} solves HSP when $G = \Z$ in
polynomial time in the length of the output. An important predecessor is
Simon's algorithm \cite{Simon:power} for the case $G = (\Z/2)^n$.  Shor's
algorithm extends to the general abelian case \cite{Kitaev:abelian}, to the
case when $H$ is normal \cite{HRT:normal}, and to the case when $H$ has few
conjugates \cite{GSVV:quantum}.  Since the main step in the generalized
algorithm is the quantum character transform on the group algebra $\C[G]$, we
will call it the \emph{character algorithm}.  

In the dihedral hidden subgroup problem (DHSP),  $G$ is the dihedral group
$D_N$ and $H$ is generated by a reflection. (Other subgroups of $D_N$ are only
easier to find; see \prop{p:arbhide}.) In this case $H$ has many conjugates and
the character algorithm works poorly. This hidden subgroup problem was first
considered by Ettinger and H{\o}yer \cite{EH:noncomm}.  They presented an
algorithm that finds $H$ with a linear number of queries (in the length of the
output) but an exponential amount of computation. Ettinger, H{\o}yer, and
Knill generalized this result to the general finite hidden subgroup problem
\cite{EHK:almost}.

In this paper we will describe a new quantum algorithm for the dihedral
group $D_N$ with a favorable compromise between query complexity 
and computation time per query.

\begin{theorem} There is a quantum algorithm that finds a hidden reflection
in the dihedral group $G = D_N$ (of order $2N$) with time and query complexity
$2^{O(\sqrt{\log\ N})}$.
\label{th:main}
\end{theorem}

The time complexity $2^{O(\sqrt{\log\ N})}$ is not polynomial, but it
is subexponential.  By contrast any classical algorithm requires at least
$2N^{1/2}$ queries on average.   Unfortunately our algorithm also requires
$2^{O(\sqrt{\log\ N})}$ quantum space.

We will prove \thm{th:main} in a convenient case, $N = 2^n$, in
\sec{s:basic}.   In \sec{s:other}, we will provide another algorithm that works
for all $N$, and we will obtain the sharper time and query complexity bound
$\tO(3^{\sqrt{2\log_3 N}})$ when $N = r^n$ for some fixed radix $r$. The
algorithm for this last case generalizes to many other smooth values of $N$.

\acknowledgments

Some elements of the algorithms in this article are due to Ettinger and
H{\o}yer \cite{EH:noncomm}.  Regev has presented some related ideas related to
lattice problems \cite{Regev:quantum}, and more recently has found a
space-efficient variation of the algorithms in this article \cite{Regev:dhsp}.
(We have also borrowed some aspects of his exposition of our algorithm.) The
author would like to thank Robert Beals, Robert Guralnick, Peter H{\o}yer, and
Eric Rains for useful discussions.  The also would also like to thank the
referees for useful comments.

\section{Group conventions}
\label{s:conven}

The dihedral group $D_N$ with $2N$ elements has the conventional presentation
$$D_N = \braket{x,y \st x^N = y^2 = yxyx = 1}.$$
(See Artin \cite[\S5.3]{Artin:algebra}.) An element of the form $x^s$ is a
\emph{rotation} and an element of the form $yx^s$ is a \emph{reflection}.  The
parameter $s$ is the \emph{slope} of the reflection $yx^s$. This terminology is
motivated by realizing $D_N$ as the symmetry group of a regular $N$-gon in the
plane (\fig{f:dihedral}).  In this model $yx^s$ is a reflection through a line
which makes an angle of $\frac{\pi s}{N}$ with the reflection line of $y$.

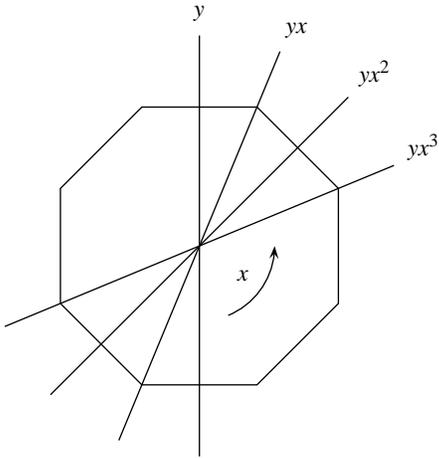
\begin{fullfigure}{f:dihedral}{Some elements of $D_8$}
\pspicture(-3.3,-3.3)(3.3,3.3)
\pspolygon(2;22.5)(2;67.5)(2;112.5)(2;157.5)
    (2;202.5)(2;247.5)(2;292.5)(2;337.5)
\psline(2.8;90)(2.8;270)
\psline(2.8;67.5)(2.8;247.5)
\psline(2.8;45)(2.8;225)
\psline(2.8;22.5)(2.8;202.5)
\rput[b](3;90){$y$}\rput[bl](3;67.5){$yx$}
\rput[bl](3;45){$yx^2$}\rput[bl](3;22.5){$yx^3$}
\psarc{->}{1}{292.5}{0}
\rput(.7;326.25){$x$}
\endpspicture
\end{fullfigure}

In this paper we will describe algorithms for the hidden subgroup problem with
$G = D_N$ and $H = \braket{yx^s}$.  If we know that the hidden subgroup is a
reflection, then the hidden subgroup problem amounts to finding its slope $s$.

\begin{proposition} Finding an arbitrary hidden subgroup $H$ of $D_N$ reduces
to finding the slope of a hidden reflection.
\label{p:arbhide}
\end{proposition}

\begin{proof} If $H$ is not a reflection, then either it is the trivial group
or it has a non-trivial intersection with the cyclic subgroup $C_N =
\braket{x}$. Finding the hidden subgroup $H' = H \cap C_N$ in $C_N$ is easy if
we know the factors of $N$, and we can factor $N$ using Shor's algorithm. 
Then the quotient group $H/H'$ is either trivial or a reflection in the
quotient group $G/H'$.

If $H$ is trivial, then this will be revealed by the fact that an algorithm to
find the slope of a hidden reflection must fail.
\end{proof}

\section{A basic algorithm}
\label{s:basic}

In this section we will describe an algorithm to find the slope $s$ of a hidden
reflection in $D_N$ when the period $N = 2^n$ is a power of 2.  The main part
of the algorithm actually only finds the parity of $s$. Once this parity is
known, the main part can be repeated with a subgroup of $D_N$ isomorphic to
$D_{N/2}$.  The group $D_N$ has two such subgroups:
$$F_0 = \braket{x^2,y} \qquad F_1 = \braket{x^2,yx}.$$
The subgroup $F_{s \bmod 2}$ contains $H$ and the other one does not, so we can
pass to one of these subgroups if and only if we know $s \bmod 2$.

For any finite set $S$, the notation $\C[S]$ denotes a Hilbert space with $S$
as an orthogonal basis.  (This is the quantum analogue of a classical data type
that takes values in $S$.)  Define the \emph{constant pure state} $\ket{S}$
in $\C[S]$, or more generally in $\C[T]$ for any $T \supseteq S$, as the 
superposition
$$\ket{S} = \frac1{\sqrt{|S|}} \sum_{s \in S} \ket{s}.$$

For the moment let us assume an arbitrary finite hidden subgroup problem $f:G
\to S$ with hidden subgroup $H$. Assuming that there is a classical circuit to
compute $f$,  we can dilate it to a unitary embedding
$$U_f:\C[G] \to \C[G] \tensor \C[S] = \C[G \times S]$$
which evaluates $f$ in the standard basis:
$$U_f\ket{g} = \ket{g,f(g)}.$$
All finite hidden subgroup algorithms, including ours, begin by computing
$$U_f\ket{G}$$
and then discarding the output register $\C[S]$, leaving the input register for
further computation.  The result is the mixed state
$$\rho_{G/H} = \frac{1}{|G|} \sum \ket{Ha}\bra{Ha}$$
on the input register $\C[G]$.

Many works on hidden subgroup algorithms describe these steps differently
\cite{Shor:factorization,NC:book,EH:noncomm,EHK:almost,GSVV:quantum,HRT:normal}. 
Instead of defining $U_f$ as an embedding that creates $f(g)$, they define it
as a unitary operator that adds $f(g)$ to an ancilla.  They describe its output
as measured rather than discarded, and they describe the mixed state
$\rho_{G/H}$ as a randomly chosen coset state $\ket{Ha}$.  We have presented an
equivalent description in the formalism of mixed states and quantum operations
\cite[Ch.8]{NC:book}.

Now let $G = D_N$ with $N = 2^n$. The general element of $D_N$ is $g = y^tx^s$
with $s \in \Z/N$ and $t \in \Z/2$.  Thus the input register $\C[D_N]$ consists
of $n$ qubits to describe $s$ and 1 qubit to describe $t$.  The second step of
our algorithm is to apply a unitary operator to $\rho_{D_N/H}$ which is almost
the character transform (\sec{s:character}). Explicitly, we apply the quantum
Fourier transform (QFT) to $\ket{s}$,
$$F_N:\ket{s} \mapsto \frac1{\sqrt{N}} \sum_k e^{2\pi i ks/N} \ket{k},$$
and then measure $k \in \Z/N$. The measured value is uniformly random, while
the state on the remaining qubit is
$$\ketpsi{k} \propto \ket{0} + e^{2\pi i ks/N} \ket{1}.$$
(The symbol ``$\propto$'' means ``proportional to'', so that we can omit
normalization and global phase.) We will always create the same state
$\rho_{D_N/H}$ and perform the same measurement, so we can suppose that we have
a supply of $2^{O(\sqrt{n}}$ states $\ketpsi{k}$, each with its own known but
random value of $k$.

Note that $\ketpsi{-k}$ and $\ketpsi{k}$ carry equivalent information about
$s$, because
\eq{e:flip}{\ketpsi{-k} = X\ketpsi{k},}
where $X$ is the bit flip operator.  They will be equivalent in our algorithms
as well.

We would like to create the state
$$\ketpsi{2^{n-1}} \propto \ket{0} + (-1)^s \ket{1}$$
because its measurement in the $\ket{\pm}$ basis reveals the parity of $s$. To
this end we create a sieve which creates new $\ketpsi{k}$'s from pairs of old
ones.  The sieve increases the number of trailing zeroes $\alpha(k)$ in the
binary expansion of $k$.  Given $\ketpsi{k}$ and $\ketpsi{\ell}$, their
joint state is
\begin{align*}
\ketpsi{k} \tensor \ketpsi{\ell} \propto
    &\ \ket{0,0} + e^{2\pi i ks/N} \ket{1,0} \\
    &\ + e^{2\pi i k\ell/N} \ket{0,1} + e^{2\pi i (k+\ell)/N} \ket{1,1}
\end{align*}
We now apply a CNOT gate
$$\ket{a,b} \mapsto \ket{a,a+b}$$
and measure the right qubit.  The left qubit has the residual state
$$\ketpsi{k \pm \ell} \propto \ket{0} + e^{2\pi i (k \pm \ell) s/N} \ket{1}$$
and the label $k \pm \ell$, which is inferred from the measurement of $a+b$. 
Thus we have a procedure to extract a new qubit $\ketpsi{k \pm \ell}$ from
the old qubits $\ketpsi{k}$ and $\ketpsi{\ell}$.  The extraction makes an
unbiased random choice between $k+\ell$ and $k-\ell$. We may well like the
extracted qubit better than either of the old ones.  

By iterating qubit extraction, we can eventually create the state that we like
best, $\ketpsi{2^{n-1}}$.  We will construct a sieve that begins with
$2^{\Theta(\sqrt{n})}$ qubits.  Each stage of the sieve will repeatedly find
two qubits $\ketpsi{k}$ and $\ketpsi{\ell}$ such that $k$ and $\ell$ agree in
$\Theta(\sqrt{n})$ low bits in addition to their trailing zeroes.  With
probability $\frac12$, the label $k \pm \ell$ of the extracted qubit has
$\sqrt{n}$ more trailing zeroes than $k$ or $\ell$.  If the sieve has depth
$\Theta(\sqrt{n})$, we can expect it to produce copies of $\ketpsi{2^{n-1}}$.

In conclusion, here is a complete description of the algorithm to find a hidden
reflection in $D_N$ with $N = 2^n$.  Also let $m = \ceil{\sqrt{n-1}}$.

\begin{algorithm} Input: An oracle $f:D_N \to S$ with a hidden
subgroup $H = \braket{yx^s}$ and $N = 2^n$.

\begin{description}
\item[1.] Make a list $L_0$ of copies of the state $\rho_{D_N/H}$ by applying
the dilation $D_f$ to the constant pure state $\ket{D_N}$ and discarding the
input. Extract $\ketpsi{k}$ from each $\rho_{D_N/H}$ with a QFT-based
measurement.

\item[2.] For each $0 \le j < m$, we assume a list $L_j$ of qubit states
$\ketpsi{k}$ such that $k$ has at least $mj$ trailing zeroes.  Divide $L_j$
into pairs of qubits $\ketpsi{k}$ and $\ketpsi{\ell}$ that share at least $m$
low bits (in addition to trailing zeroes), or $n-1-mj$ bits if $m=j-1$. 
Extract the state $\ketpsi{k \pm \ell}$ from each pair.  Let the new list
$L_{j+1}$ consist of those qubit states of the form $\ketpsi{k - \ell}$.

\item[3.] The final list $L_m$ consists of states $\ketpsi{0}$ and
$\ketpsi{2^{n-1}}$.  Measure a state $\ketpsi{2^{n-1}}$ in the $\ket{\pm}$
basis to determine the parity of the slope $s$.

\item[4.] Repeat steps 1-3 with the subgroup of $D_N$ which is isomorphic to
$D_{N/2}$ and which contains $H$.

\end{description} \label{a:stages}
\end{algorithm}

\subsection{Proof of the complexity}

\begin{theorem} \alg{a:stages} requires $O(8^{\sqrt{n}})$ queries
and $\tO(8^{\sqrt{n}})$ computation time.
\label{th:stages}
\end{theorem}

\begin{proof} In outline, if $|L_j| \gg 2^m$, then we can pair almost all of
the elements of $L_j$ so that $k$ and $\ell$ share $m$ low bits for
each pair $\ketpsi{k}$ and $\ketpsi{\ell}$.  Then about half of the
pairs will form $L_{j+1}$, so that
$$\frac{|L_{j+1}|}{|L_j|} \approx \frac14.$$
We can set $|L_m| = \Theta(2^m)$.  Working backwards,
we can set $|L_0| = \Theta(8^m)$.  The computation time consists of 
tasks with only logarithmic overhead.

In detail, we will assume that
$$|L_j| \ge C_{m-j} 2^{3m-2j}$$
for a certain constant $9 > C_k \ge 3$.  We will bound the probability that
this assumption survives as $j$ increases. The constants are defined by letting
$C_0 = 3$, and letting
$$C_k = \frac{C_{k-1}}{1-2^{-k-\frac{m}3}} + 2^{-2k}$$
by induction on $k$.  It is not hard to check that
$$C_k > C_{k-1} \qquad \lim_{k \to \infty} C_k < 9.$$
(A calculator may help for the first few terms of the limit, the worst case
being $m=1$.)

Since we create $L_0$ directly from oracle calls, we can set
$$|L_0| = C_0 2^{3m}.$$
Given $L_j$, let $P_j$ be a maximal set of pairs $\ketpsi{k}$ and
$\ketpsi{\ell}$ with $m$ low matching bits. Then
$$|P_j| \ge \frac{|L_j|-2^m}{2} \ge \frac{2^{3m-2j}C_j(1-2^{2j-2m})}{2},$$
because there are at most $2^m$ unmatched pairs.  The list $L_{j+1}$ is then
formed from $P_j$ by summand extraction, so $|L_{j+1}|$ can be understood as
the sum of $N$ independent, unbiased Bernoulli random variables. In general if
$B_N$ is a sum of $N$ unbiased Bernoulli random variables, then
$$P[B_N \le \frac{(1-b)N}{2}] \le (\cosh b)^Ne^{-Nb^2} \le e^{-Nb^2/2}.$$
(The first inequality is the Chernoff bound on large deviations.)
Setting 
$$b = 2^{j-\frac{4m}{3}},$$
we learn that
$$|L_{j+1}| \ge \frac{2^{3m-2j}(C_j-2^{2j-2m})(1-2^{j-\frac{4m}{3}})}{4} =
    C_{j+1}2^{3m-2j-2}$$
with probability at least
$$1-e^{-2^{\frac{m}3-1}}.$$
Finally by induction on $j$,
$$P[|P_j| \ge C_{m-j} 2^{3m-2j}\;\forall j]
    \ge (1-e^{-2^{\frac{m}3-1}})^m \to 1$$
as $m \to \infty$.  

Thus the final list $L_m$ is very likely to be large.  Since the highest bit of
$k$ in $\ketpsi{k}$ was never used for any decisions in the algorithm, it is
unbiased Bernoulli for each entry of $L_m$. Therefore $L_m$ is very likely to
contain copies of $\ketpsi{2^{n-1}}$.
\end{proof}

\section{Some motivation}
\label{s:motivation}

\alg{a:stages} can be motivated by related ideas in representation theory and
the theory of classical algorithms.

On the representation theory side, the input space $\C[D_N]$ has an orthogonal
decomposition into 2-dimensional representations $V_k$ of $D_N$,
\eq{e:vdecomp}{\C[D_N] \cong \bigoplus_{k \in \Z/N} V_k.}
This means that each element of $D_N$ is represented by a unitary operator on
$\C[D_N]$ (given by left multiplication) and each $V_k$ is an invariant
subspace, so that each element of $D_N$ is also represented by a unitary
operator on each $V_k$ \cite[\S9.2]{Artin:algebra}.  Every orthogonal
decomposition of a Hilbert space corresponds to a projective measurement
\cite[\S2.2.5]{NC:book}; this particular measurement can be computed using a
QFT.

In the representation $V_k$, the generators $x$ and $y$ are
represented as follows:
$$x \mapsto \begin{pmatrix} e^{2\pi/N} & 0 \\ 0 & e^{-2\pi/N}
    \end{pmatrix} \qquad y \mapsto
    \begin{pmatrix} 0 & 1 \\ 1 & 0 \end{pmatrix}.$$
Since the state $\ket{Ha}$ is invariant under the represented action of $H$,
the residual state $\ketpsi{k}$ is too.  Thus abstract representation theory
motivates the use of this state to find $H$. Note also that $V_k \cong V_{-k}$
as representations, as if reflected in the equivalence between $\ketpsi{k}$ and
$\ketpsi{-k}$ in equation~\eqref{e:flip}.

The representation $V_k$ is irreducible except when $k = 0$ or $k = N/2$.  Thus
equation~\eqref{e:vdecomp} is not far from the Burnside decomposition of
$\C[G]$ into irreducible representations in the special case $G = D_N$. When
expressed as a unitary operator, the Burnside decomposition is called the
\emph{character transform} or the non-commutative Fourier transform.  (Measuring the
character name solves the hidden subgroup problem for normal subgroups
\cite{HRT:normal} and almost normal subgroups \cite{GSVV:quantum}.)  Using
$V_{N/2}$ as the target of \alg{a:stages} is motivated by its reducibility; the
measurement corresponding to its irreducible decomposition is the one that
reveals the slope of $s$.

On the algorithm side, the sieve in \alg{a:stages} is similar to a sieve algorithm
for a learning problem due to Blum, Kalai, and Wasserman \cite{BKW:learning}
and to a sieve to find shortest vector in a lattice due to Ajtai, Kumar, and
Sivakumar \cite{AKS:lattice}.

Ettinger and H{\o}yer \cite{EH:noncomm} observed that if the state $\ketpsi{k}$
for the hidden subgroup $H = \braket{x^sy}$ will be found in the state
$\ket{\psi'_k}$ for a reference subgroup $H' = \braket{x^ty}$ with probability
$$\cos(\pi i(s-t)k/N)^2.$$
Thus the state $\ketpsi{k}$ can provide a coin flip with this bias.  We call
such a coin flip a \emph{cosine observation} of the slope $s$. Ettinger and
H{\o}yer showed that $s$ is revealed by a maximum likelihood test with respect
to $O(\log\ N)$ cosine observations with random values of $k$.  They suggested
a brute-force search to solve this maximum likelihood problem.  Our first
version of \alg{a:stages} was a slightly subexponential, classical sieve on cosine
observations that even more closely resembles the Blum-Kalai-Wasserman
algorithm. Replacing the cosine observations by the qubit states $\ketpsi{k}$
themselves significantly accelerates the algorithm.

\section{Other algorithms}
\label{s:other}

\alg{a:stages} presents a simplified sieve which is close to the author's
original thinking.  But it is neither optimal nor fully general.  In this
section we present several variations which are faster or more general.

The first task is to prove \thm{th:main} when $N$ is not a power of 2.   Given
any qubit state $\ketpsi{k}$, we can assume that $0 \le k \le \frac{N}2$, since
$\ketpsi{k}$ and $\ketpsi{-k}$ are equivalent.  The list $L_j$
will consist of qubits $\ketpsi{k}$ with
$$0 \le k < 2^{m^2-mj+1},$$
where
$$m = \bceil{\sqrt{(\log_2 N)-2}}.$$

Another difference when $N$ is not a power of 2 is that the quantum Fourier
transform on $\Z/N$ is more complicated.  An efficient approximate algorithm
was given by Kitaev \cite{Kitaev:abelian}; another algorithm which is exact (in
a sense) is due to Mosca and Zalka \cite{MZ:exact}.

\begin{algorithm}  Input: An oracle $f:D_N \to S$ with a hidden
subgroup $H = \braket{yx^s}$.

\begin{description}
\item[1.] Make a list $L_0$ of copies of $\rho_{D_N/H}$. Extract a qubit state
$\ketpsi{k}$ from each $\rho_{D_N/H}$ using a QFT on $\Z/N$ and a measurement.

\item[2.] For each $0 \le j < m$, we assume a list $L_j$ of qubit states
$\ketpsi{k}$ such that $0 \le k \le 2^{m^2-mj+1}$.  Randomly divide $L_j$
into pairs of qubits $\ketpsi{k}$ and $\ketpsi{\ell}$ that
such that
$$|k-\ell| \le 2^{m^2-m(j+1)+1}.$$
Let the new list $L_{j+1}$ consist of those qubit states of the form
$\ketpsi{|k - \ell|}$.


\item[3.] The final list $L_m$ consists of states $\ketpsi{0}$ and
$\ketpsi{1}$.  Perform the Ettinger-H{\o}yer measurement on the copies of
$\ketpsi{1}$ with different values of $t$ to learn $s \in \Z/N$ to within
$N/4$.

\item[4.] Write $N = 2^aM$ with $M$ odd.  By the Chinese remainder theorem,
$$C_N \cong C_{2^a} \times C_M.$$
For each $1 \le j \le \lceil \log_2 N \rceil$, apply \alg{a:stages} to produce
many $\ketpsi{k}$ with $2^{\min(a,j)}|k$. Then repeat steps 1-4 after applying
the group automorphism $x \mapsto x^{2^{-j}}$ to the $C_M$ factor of $D_N$. 
This produces copies of $\ketpsi{2^j}$, hence cosine observations $\cos(\pi
i2^j(s-t)/N)^2$.  These observations determine $s$.
\end{description}
\label{a:high} \end{algorithm}

The proof of \thm{th:stages} carries over to show that \alg{a:high} also only
requires $O(8^{\sqrt{\log_2 N}})$ queries, and quasilinear time in its data. 
The only new step is to check that in the final list $L_m$, the qubit states
$\ketpsi{0}$ and $\ketpsi{1}$ are almost equally likely.  This is  a bit
tricky, but inevitable given that the lowest bit of $k$ can be almost
uncorrelated with the way that $\ketpsi{k}$ is paired.

\begin{remark} Peter H{\o}yer describes a simplification of \alg{a:high}
\cite{Hoyer:personal}.  Given only one copy each of
$$\ketpsi{1},\ketpsi{2},\ldots,\ketpsi{2^k},$$
with $2^k \ge N$, the slope $s$ can be recovered directly by a quantum Fourier
transform.  More precisely, the measured Fourier number $t$ of these qubits
reveals $s$ by the relation
$$\frac{t}{2^k} \sim \frac{s}{N}.$$
This simplification saves a factor of $O(\log N)$ computation time.
\end{remark}

Now suppose that $N = r^n$ for some small radix $r$; \alg{a:stages} generalizes
to this case with only slight changes. It is natural to accelerate it by
recasting it as a greedy algorithm.  To this end, we define an objective
function $\alpha(k)$ that expresses how much we like a given state
$\ketpsi{k}$.  Namely, let $\alpha{k}$ be the number of factors of $r$ in $k$,
with the exception that $\alpha(0) = 0$.  Within the list $L$ of qubit states
available at any given time, we will greedily pick $\ketpsi{k}$ and
$\ketpsi{\ell}$ to maximize $\alpha(k \pm \ell)$.  It is also natural to
restrict our greed to the qubits that minimize $\alpha$, because there is no
advantage to postponing their use in the sieve.

\begin{algorithm} Input: An oracle $f:D_N \to S$ with a hidden
subgroup $H = \braket{yx^s}$ and $N = r^n$.

\begin{description}
\item[1.] Make a list $L$ of qubit states $\ketpsi{k}$
extracted from copies of $\rho_{D_N/H}$.

\item[2.] Within the sublist $L'$ of $L$ that minimizes $\alpha$, repeatedly
extract $\ketpsi(k \pm \ell)$ from a pair of qubits $\ketpsi{k}$ and
$\ketpsi{\ell}$ that maximize $\alpha(k \pm \ell)$.

\item[3.] After enough qubits $\ketpsi{k}$ appear with $\frac{N}r | k$,
measure $s \mod r$ using state tomography.  Then repeat the algorithm with a
subgroup of $D_N$ isomorphic to $D_{N/r}$.
\end{description}
\label{a:low} \end{algorithm}

The behavior of \alg{a:low} (but not its quantum state) can be simulated by a
classical randomized algorithm.   We include the source code of a simulator
written in Python with this article \cite{Kuperberg:dhspsim.py} with $r=2$. Our
experiments with this simulator led to a false conjecture for algorithm's
precise query complexity.  Nonetheless we present some of its results in
\tab{t:bits}.  The last line of the table are roughly consistent with
\thm{th:3}.  Note that the sieve is a bit more efficient when $r=2$
because then $k \pm \ell$ increases by 1 in the unfavorable case
and at least 2 in the favorable case.

\begin{fulltable}{t:bits}{r|rrrrrrrr}
    {Average cancelled bits in a simulation (100 trials)}
Queries & 3 & $3^2$ & $3^3$ & $3^4$ & $3^5$ & $3^6$ & $3^7$ & $3^8$ \\ \hline
Zeroed bits & 3.62 & 6.75 & 12.53 & 19.07 & 27.14 & 36.44 & 47.51 & 59.76 \\
$\sqrt{2\log_3 2^n}$ & 2.14 & 2.92 & 3.98 & 4.91 & 5.85 & 6.78 & 7.74 & 8.68
\end{fulltable}

\begin{theorem} \alg{a:low} requires $\tO(3^{\sqrt{2\log_3 N}})$
queries and quasilinear time in the number of queries.
\label{th:3} \end{theorem}

Here is a heuristic justification of the query bound in \thm{th:3}. We assume,
as the proof will, that $r=3$ and $N = 3^n$.  Then with $3^{\sqrt{2n}}$
queries, we can expect qubit extraction to initially cancel about $\sqrt{2n}$
ternary digits (trits) with probability $\frac12$.  If we believe the query
estimate for $n' < n$, then we can expect the new qubit to be about 3 times as
valuable as the old one, since
$$\sqrt{2n} - \sqrt{2n - \sqrt{2n}} \approx 1.$$
Such a qubit extraction trades 2 qubits for 1 qubit which is half the time
equivalent to the original 2 and half the time 3 times as valuable. Thus each
step of the sieve breaks even; it is like a gamble with \$2 that is equally
likely to return \$1 or \$3.

\begin{proof} (Sketch) We will show that the sieve produces states
$\ketpsi{aN/r}$ (which we will call \emph{final states}) with adequate
probability when provided with at least $C n 3^{\sqrt{2\log_3 N}}$ queries. The
work per query is quasilinear in $|L|$ (initially the number of queries) if the
list $L$ is dynamically sorted.  To simplify the formulas, we assume that
$r=3$, although the proof works for all $r$.

We can think of a qubit state $\ketpsi{k}$ as a monetary asset, valued by
the function
$$V(k) = 3^{-\sqrt{2(n-1-\alpha(k))}}.$$
Thus the total value $V(L)$ of the initial list $L$ is at least
$$V(L) \ge Cn.$$

We claim that over a period of the sieve that increases $\min \alpha$ by 1, the
expected change in $V(L)$ is at worst $-C$.  Since $\min \alpha$ can only
increase $n-1$ times, $V(L) \ge C$ when $\min \alpha = n-1$.  Thus the sieve
produces at least $C$ final states on average.  Along the way, the changes to
$V(L)$ are independent (but not identically distributed) Bernoulli trials.  One
can show using a version of the Chernoff bound (as in the proof of
\thm{th:stages}) that the number of final states is not maldistributed.  We
will omit this refinement of the estimates and spell out the expected behavior
of $V(L)$.

Given $k$, let
$$\beta = \beta(k) = n-1-\alpha(k)$$
for short, so that $\beta$ can be thought of as the number of
uncancelled trits in the label $k$ of $\ketpsi{k}$.
Suppose that two labels $k$ and $\ell$ or $-\ell$ share $m$ trits in addition
to $\alpha(k)$ cancelled trits. Then
\eq{e:value}{V(k) = V(\ell) = 3^{-\sqrt{2\beta}}.}
The state $\ketpsi{k \pm \ell}$ extracted from $\ketpsi{k}$ and $\ketpsi{\ell}$
has the expected value
\begin{align}
E[V(k \pm \ell)] &= \frac{3^{-\sqrt{2\beta}}
    + 3^{-\sqrt{2(\beta-m)}}}2 \nonumber \\
    &> 2V(k) \frac{1+3^{m/\sqrt{2\beta}}}4, \label{e:gamble}
\end{align}
using the elementary relation
$$\sqrt{2\beta} - \sqrt{2(\beta-m)}
    = \frac{2m}{\sqrt{2\beta} + \sqrt{2(\beta-m)}}
    > \frac{m}{\sqrt{2\beta}}.$$

The most important feature of equation~\eqref{e:gamble} is that if $m >
\sqrt{2\beta}$, the expected change in $V(L)$ is positive.  Thus in bounding
the attrition of $V(L)$, we can assume that $m \le \sqrt{2\beta}$ for the
best-matching qubits $\ketpsi{k}$ and $\ketpsi{\ell}$ in the sublist $L'$ that
minimizes $\alpha$. By the pigeonhole principle, this can only happen when
$$|L'| \le 3^{\sqrt{2\beta}}.$$
(To apply the pigeonhole principle properly, use the equivalence between
$\ketpsi{k}$ and $\ketpsi{-k}$ to assume that the first non-zero digit is 1.
There are then $3^m$ choices for the next $m$ digits.) 

When qubit extraction decreases $V(L)$, it decreases by at worst the value of
one parent, given by the right side of \eqref{e:value}. Likewise if $|L'| = 1$
and its unique element $\ketpsi{k}$ must be discarded, the loss to $V(L)$ is
again the right side of \eqref{e:value}. Thus the total expected loss as $L'$
is exhausted is at most
$$3^{-\sqrt{2\beta}} 3^{\sqrt{2\beta}} < 1.$$
We can therefore take $C=1$, although a larger $C$ may be convenient to
facilitate the Chernoff bound.
\end{proof}

\begin{remark} A close examination of \alg{a:low} and \thm{th:3} reveals
that the sieve works with the same complexity bound if $N$ factors as
$$N = N_1 N_2 \ldots N_m$$
and $N_k$ is within a bounded factor of $3^k$.  In this case
the sieve will determine $s \bmod N_1$.  This is enough values
of $N$ to extend to an algorithm for all $N$ by the
method of spliced approximation \sec{s:more}.
\end{remark}

\section{Generalized dihedral groups and hidden shifts}
\label{s:shifts}

In this section we consider several other problems that are equivalent or
closely related to the hidden dihedral subgroup problem.

In general if $A$ is an abelian group, let $\exp(A)$ denote the multiplicative
form of the same group.  Let $C_n = \exp(\Z/n)$ be the multiplicative cyclic
group of order $n$.  If $A$ is any abelian group, define the \emph{generalized
dihedral group} to be the semidirect product
$$D_A \cong C_2 \ltimes \exp(A)$$
with the conjugation relation
$$x^{-1} = yxy$$
for all $x \in \exp(A)$ and for the non-trivial $y \in C_2$.
Any element of the form $yx$ is a \emph{reflection} in $D_A$.

Suppose that $A$ is an abelian group and $f,g:A \to S$ are two injective
functions that differ by a shift:
$$f(a) = g(a+s).$$
Then the task of finding $s$ from $f$ and $g$ is the \emph{abelian
hidden shift problem}.  Another problem is the hidden reflection
problem in $A$ (as opposed to in $D_A$).  In this problem,
$f:A \to S$ is a function which is injective except that
$$f(a) = f(s-a)$$
for some hidden $s$.

\begin{proposition} If $A$ is an abelian group, the hidden shift and hidden
reflection problems in $A$ are equivalent to the hidden reflection problem in
$D_A$.
\end{proposition}

See \tab{t:shift} for an example.

\begin{fulltable}{t:shift}{r|cccccccc}
    {An oracle that hides $\braket{yx^3}$ in $D_8$ and its hidden shift}
$a$ & $1$ & $x$ & $x^2$ & $x^3$ & $x^4$ & $x^5$ & $x^6$ & $x^7$ \\ \hline
$f(a)$ & A & B & C & D & E & F & G & H \\ \hline \hline
$a$ & $y$ & $yx$ & $yx^2$ & $yx^3$ & $yx^4$ & $yx^5$ & $yx^6$ & $yx^7$ \\ \hline
$f(a)$ & F & G & H & A & B & C & D & E \\
\end{fulltable}

\begin{proof} If $a \in A$, let $x^a$ denote the corresponding
element in $\exp(A)$.  Given $f,g:A \to S$, define
$$h(x^a) = f(a) \qquad h(yx^a) = g(a).$$
Then evidently
$$h(x^a) = h(yx^{s+a})$$
if and only if
$$f(a) = g(a+s).$$

We can also reduce the pair $f$ and $g$ to a function with a hidden
reflection.  Namely let $S^{(2)}$ be the set of unordered pairs of
elements of $S$ and define $h:A \to S^{(2)}$ by
$$h(a) = \{f(-a),g(a)\}.$$
Then $h$ is injective save for the relation
$$h(a) = h(s-a).$$

Contrariwise suppose that $h:A \to S$ is injective save for the relation
$$h(a) = h(s-a).$$
If there is a $v \in A$ such that $2v \ne 0$, define
$$f:A \to S^{\times 2} \qquad g:A \to S^{\times 2}$$
by 
$$f(a) = (h(-a),h(v-a)) \qquad g(a) = (h(a),h(a-v)).$$
(If $A$ is cyclic, we can just take $v=1$.) Then $f$ and $g$ are injective and
$$f(a) = g(a+s).$$
If all $v \in A$ satisfy $2v = 0$, then $h$ hides a subgroup
of $A$ generated by $s$, so we can find $s$ by Simon's algorithm.
\end{proof}

Note also that \prop{p:arbhide} generalizes readily to generalized dihedral
subgroups: finding a hidden reflection in $D_A$ is as difficult as finding any
hidden subgroup.

A final variation of DHSP is the hidden substring problem.
In the $N \injects M$ hidden substring problem,
\begin{align*}
f:\{0,1,2,\ldots,N-1\} &\to S \\
g:\{0,1,2,\ldots,M-1\} &\to S
\end{align*}
are two injective functions such that $f$ is a shifted restriction of $g$, \ie,
$$f(x) = g(x + s)$$
for all $0 \le x < N$ and for some fixed $0 \le s < M-N$.

\section{More algorithms}
\label{s:more}

In this section we will establish a generalization of \thm{th:main}
and a corollary:

\begin{theorem} The abelian hidden shift problem has an algorithm
with time and query complexity $2^{O(\sqrt{n})}$, where $n$ is the length of
the output, uniformly for all finitely generated abelian groups.
\label{th:abelian} \end{theorem}

\begin{corollary} The $N \injects 2N$ hidden substring problem has an algorithm
with time and query complexity $2^{O(\sqrt{\log\ N})}$.
\label{c:substring} \end{corollary}

The proof of \cor{c:substring} serves as a warm-up to the proof of
\thm{th:abelian}.  It introduces a technique for converting hidden shift
algorithms that we call \emph{spliced approximation}.

\begin{proof}[Proof of \cor{c:substring}]
Identify the domain of $f$ with $\Z/N$.  (No matter that this
identification is artificial.)  Make a random estimate $t$ for the value
of $s$, and define $h:D_N \to S$ by
$$g'(n) = g(n + t).$$
If $t$ is a good estimate for $s$, then $f$ and $g'$ approximately hide the
hidden shift $s-t$.  If we convert $f$ and $g$ to a function $h:D_N \to S$,
then apply its dilation $U_h$ with input $\ket{D_N}$ and discard the output,
the result is a state $\rho_h = \rho_{f,g'}$ which is close to the state
$\rho_{D_N/H}$ used in \alg{a:high}.

We need to quantify how close.  The relevant metric on states for us is the
trace distance \cite[\S9.2]{NC:book}.  In general if $\rho$ and $\rho'$ are two
states on a Hilbert space $\cH$, the trace distance $||\rho - \rho'||$ is the
maximum probability that any measurement, indeed any use in a quantum
algorithm, will distinguish them.  In our case,
$$||\rho_h - \rho_{D_N/H}|| = \frac{|s-t|}{N}.$$
If
$$\frac{|s-t|}{N} = 2^{-O(\sqrt{\log\ N})},$$
then with bounded probability, \alg{a:high} will never see the difference
between $\rho_h$ and $\rho_{D_N/H}$. Thus $2^{O(\sqrt{\log\ N})}$ guesses for
$s$ suffice.
\end{proof}

A second warm-up to the general case of \thm{th:abelian} is the special case $A
= \Z$.  Recall that more computation is allowed for longer output.  Suppose
that the output has $n$ bits, \ie, the shift $s$ is at most $2^n$.  In the
language of deterministic hiding, we restrict the domain of $f,g:\Z \to S$ to
the set $\{0,1,2,\ldots,2^m\}$, where $m = n + \Theta(\sqrt{n})$, and interpret
this set as $\Z/2^m$.  Then $f$ and $g$ approximately differ by the shift $s$. 
If we form the state $\rho_{f,g}$ as in the proof of \cor{c:substring}, then
its trace distance from the state $\rho_{D_N/H}$, with $N = 2^m$, is
$2^{-O(\sqrt{n})}$.  Thus \alg{a:high} will never see the states differ.

\begin{proof}[Sketched proof of \thm{th:abelian}]
In the general case, the classification of finitely generated abelian groups
says that
$$A \cong \Z^b \oplus \Z/N_1 \oplus \Z/N_2 \oplus \cdots \oplus \Z/N_a.$$
Assuming a bound on the length of the output, we can truncate each $\Z$ summand
of $A$, as in the case $A = \Z$.  (We suppose that we know how many bits of
output are allocated to each free summand of $A$.) Thus we can assume that
$$A = \Z/N_1 \oplus \Z/N_2 \oplus \cdots \oplus \Z/N_a,$$
and the problem is to find $s$ in time $2^{O(\sqrt{\log\ |A|})}$. In other
words the problem is to solve HSP for a finite group $D_A$.

The general element of $D_A$ can be written $y^tx^a$ with $t \in \Z/2$ and $a
\in A$.  Following the usual first step, we can first prepare the state
$\rho_{D_A/H}$.  Then we can perform a quantum Fourier transform on each factor
of $A$, then measure the answer, to obtain a label
$$k = (k_1,k_2,\ldots,k_a) \in A$$
and a qubit state
$$\ketpsi{k} \propto \ket{0} + e^{2\pi i \sum_j s_jk_j/N_j} \ket{1}.$$
(As in \sec{s:motivation}, this state is $H$-invariant in a
two-dimensional representation $V_k$ of $D_A$.)  We will outline
a sieve algorithm to compute any one coordinate of the slope,
without loss of generality $s_a$.

As in \alg{a:low}, we will guide the behavior of the sieve by an objective
function $\alpha$ on $A$.  Given $k$, let $b(k)$ be the first $j$ such that
$k_j \ne 0$. If $b<a$, then let
$$\alpha(k) = \sum_{j=1}^b \lceil 1+\log_2 (N_j+1) \rceil
    - \lceil \log_2 (k_b+1) \rceil.$$
If $b = a$, then let
$$\alpha(k) = \sum_{j=1}^a \lceil 1+\log_2 (N_j+1) \rceil.$$
As in \alg{a:low}, we produce a list $L$ of $2^{O(\sqrt{\log\ |A|})}$ qubits
with states $\ketpsi{k}$.  Within the minimum of $\alpha$ on $L$, we
repeatedly find pairs $\ketpsi{k}$ and $\ketpsi{\ell}$ that maximize
$\alpha(k+\ell)$ or $\alpha(k-\ell)$, then we extract $\ketpsi{k+\ell}$ from
each such pair.  The end result is a list of qubit states $\ketpsi{k}$ with
$$k = (0,0,\ldots,0,k_a).$$
The set of $k$ of this form is closed under sums and differences, so we can
switch to \alg{a:high} to eventually determine the slope $s_a$.
\end{proof}

Note that many abelian groups $A$ are not very different from cyclic groups, so
that the generalized dihedral group $D_A$ can be approximated for our purposes
by a standard dihedral group. For example, if $A \cong \Z^a$ is free abelian
with many bits of output allocated to each coordinate, then we can pass to a
truncation $$\Z/N_1 \oplus \Z/N_2 \oplus \cdots \oplus \Z/N_a$$ with relatively
prime $N_j$'s.  In this case the truncation is cyclic.

\section{Hidden subgroup generalities}

In this section we will make some general observations about quantum algorithms
for hidden subgroup problems.  Our comments are related to work by Hallgren,
Russell, and Ta-Shma \cite{HRT:normal} and by Grigni, Schulman, Vazirani, and
Vazirani \cite{GSVV:quantum}. 

\subsection{Quantum oracles}
\label{s:oracles}

The first step of all quantum algorithms for the hidden subgroup problem is to
form the state $\rho_{G/H}$, or an approximation when $G$ is infinite, except
when the oracle $f:G \to S$ has special properties.

Suppose that a function $f:G \to S$ that hides the subgroup $H$.  We can say
that  $f$ \emph{deterministically} hides $H$ because it is a deterministic
function.  Some problems in quantum computation might reduce to a
non-deterministic oracle $f:G \to \cH$, where $\cH$ is a Hilbert space.  We say
that such an $f$ \emph{orthogonally} hides $H$ if $f$ is constant on each right
coset $Ha$ of $H$ and orthogonal on distinct cosets.  If a quantum algorithm
invokes the dilation $D_f$ of $f$ and then discards the output, then it solves
the orthogonal hidden subgroup problem as well as the deterministic one.

Computing $D_f$ and discarding its output can also be viewed as a quantum
oracle.  A general quantum computation involving both unitary and non-unitary
actions can be expressed as a quantum operation \cite[Ch.8]{NC:book}.  In this
case the operation is a map $\cE_{G/H}$ on $\cM(\C[G])$, where in general
$\cM(\cH)$ denotes the algebra of operators on a Hilbert space $\cH$.  It is
defined by
$$\cE_{G/H}(\ket{a}\bra{b}) =
\begin{cases} \ket{a}\bra{b} & \text{if $Ha = Hb$} \\
    0 & \text{if $Ha \ne Hb$} \end{cases}.$$
We say that the quantum oracle $\cE_{G/H}$ \emph{projectively} hides the
subgroup $H$.  Unlike deterministic and orthogonal oracles, the projective
oracle is uniquely determined by $H$.  Again, all quantum algorithms for hidden
subgroup problems work with this more difficult oracle.

Finally if $G$ is finite, the projective oracle $\cE_{G/H}$ can be applied to
the constant pure state $\ket{G}$ to produce the state
$$\rho_{G/H} = \frac{|H|}{|G|} \sum \ket{Ha}\bra{Ha}.$$
So an algorithm could use a no-input oracle that simply broadcasts copies of
$\rho_{G/H}$.  Such an oracle \emph{coherently} hides $H$.  This oracle has
been also been called the random coset oracle \cite{Regev:quantum} because the
state $\rho_{G/H}$ is equivalent to the constant pure state $\ket{Ha}$ on a
uniformly randomly chosen coset.  Almost all existing quantum algorithms for
finite hidden subgroup problems only need copies of the state $\rho_{G/H}$.
\alg{a:stages} and \alg{a:low} are exceptions:  They use $\rho_{D_N/H}$ to find
the parity of the slope $s$, then relies on $\cE_{D_N/H}$ with other inputs
(constant pure states on subgroups) for later stages.  The possibly slower
algorithm \alg{a:high} works with the coherent oracle; it uses only
$\rho_{D_N/H}$.

The distinctions between deterministic, orthogonal, and projective hiding apply
to any hidden partition problem.  In one special case, called the hidden
stabilizer problem \cite{Kitaev:abelian}, a group $G$ acts transitively on a
set $S$ and a function $f:S \to T$ is invariant under a subgroup $H \subseteq
G$.  The hidden stabilizer problem has enough symmetry to justify consideration
of coherent hiding. It would be interesting to determine when one kind of
hiding is harder than another. For example, if $f$ is injective save for a
single repeated value, then there is a sublinear algorithm for deterministic
hiding \cite{BDHHMSdW:distinctness}.  But projective hiding requires at least
linear time and we do not know an algorithm for coherent hiding which is faster
than quadratic time.

In a variant of coherent HSP, the oracle outputs non-uniform mixtures of coset
states $\ket{Ha}$.  The mixtures may even be chosen adversarially.  This can
make the subgroup $H$ less hidden, for example in the trivial extreme in which
the state is $\ket{H}$ with certainty.  At the other extreme, we can always
uniformize the state by translating by a random group element.  Thus uniform
coherent HSP is the hardest representative of this class of problems.

\subsection{The character measurement}
\label{s:character}

The second step of all quantum algorithms for the generic hidden subgroup
problem is to perform the character measurement.  (The measurement in our
algorithms is only trivially different.) The result is the name or character of
an irreducible unitary representation (or irrep) $V$ and a state in $V$.
Mathematically the character measurement is expressed by the Burnside
decomposition of the group algebra $\C[G]$ as a direct sum of matrix algebras
\cite{Serre:finite}:
$$\C[G] \cong \bigoplus_V \cM(V).$$
Here $\cM(V)$ is the algebra of operators on the irrep $V$; the direct sum runs
over one representative of each isomorphism type of unitary irreps. The group
algebra $\C[G]$ has two commuting actions of $G$, given by left and right
multiplication, and with respect to these two actions,
$$\cM(V) \cong V \tensor V^*,$$
so that the Burnside decomposition can also be written
\eq{e:burnside}{\C[G] \cong \bigoplus_V V \tensor V^*.}
In light of the identification with matrices, the factor of $V^*$ is called the
\emph{row space}, while the factor of $V$ is the \emph{column space}.

The Burnside decomposition is also an orthogonal decomposition of Hilbert
spaces, and so corresponds to a projective measurement on $\C[G]$.  This is the
character measurement.  A character transform is an orthonormal change of basis
that refines equation \eqref{e:burnside}.  Its precise structure as a unitary
operator depends on choosing a basis for each $V$.

The state $\rho_{G/H}$ has an interesting structure with respect to
the Burnside decomposition. In general if $\cH$ is a finite-dimensional Hilbert
space, let $\rho_{\cH}$ denote the uniform mixed state on $\cH$; while if $V$
is a representation of a group $G$, let $V^G$ denote its invariant space.  It
is easy to check that 
$$\rho_{G/H} = \rho_{\C[G]^H},$$
where $G$ (and therefore $H$) acts on $\C[G]$ by left multiplication. In the
Burnside decomposition, the left multiplication action on each $V \tensor V^*$
is trivial on the right factor $V^*$ and is just the defining action of $G$ on
$V$.  Since $\rho_{G/H}$ is the uniform state on all $H$-invariant vectors in
$\C[G]$, this property descends through the Burnside decomposition:
$$\rho_{G/H} = \bigoplus_V \rho_{V^H} \tensor \rho_{V^*}.$$
This relation has two consequences.  First, as has been noted previously
\cite{GSVV:quantum}, the state on the row space $V^*$ has no useful
information.  Second, since $\rho_{G/H}$ decomposes as a direct sum with
respect to the Burnside decomposition, the character measurement sacrifices no
coherence to the environment; it only measures something that the environment
already knows.  Our reasoning here establishes the following proposition:

\begin{proposition} Let $G$ be a finite group and assume an algorithm or oracle
to compute the character transform on $\C[G]$. Then a process provides the
state $\rho_{G/H}$ is \emph{equivalent} to a process that provides the name of
an irrep $V$ and the state $\rho_{V^H}$ with probability
$$P[V] = \frac{(\dim V)(\dim V^H)|H|}{|G|}.$$
\label{p:char}
\end{proposition}

\prop{p:char} sharpens the motivation to work with irreps in the hidden
subgroup problem.  If you obtain the state $\rho_{G/H}$, and if you can
efficiently perform the character measurement on states, then you might as well
apply it to $\rho_{G/H}$.

\prop{p:char} and the definition of coherent HSP in \sec{s:oracles} suggest
another class of oracles related to the hidden subgroup problem.  In general an
oracle might provide the name of a representation $V$ and a state $\rho$ which
is some mixture of $H$-invariant pure states in $V$.  It is tempting to
describe such a $\rho$ as $H$-invariant, but technically that is a weaker
condition that also applies to other states.
For example, the uniform state on $V$ is $H$-invariant.  So we
say that $\rho$ is \emph{purely $H$-invariant} if it is supported on in the
$H$-invariant space $V^H$.  For example, the uniform state $\rho_{G/H}$ is
purely $H$-invariant.  More generally the purely $H$-invariant state on
$\C[G]$ are exactly the mixtures of constant pure states of right cosets
$\ket{Ha}$.

\begin{proposition} Let $G$ be a finite group.  Then any purely $H$-invariant
state $\rho$ on $\C[G]$ can be converted to $\rho_{G/H}$.  In the presence of
an algorithm or oracle to perform the character transform on $\C[G]$, any
purely $H$-invariant state $\rho$ on any irrep $V$ can be converted to
$\rho_{G/H}$.
\label{p:worst} \end{proposition}
\begin{proof} If we right-multiply $\rho$ by a uniformly random element of $G$,
it becomes $\rho_{G/H}$.  If we perform the reverse character transform 
to a purely $H$-invariant state $\rho$ on $V$, it becomes a purely
$H$-invariant state on $\rho_{G/H}$ itself.
\end{proof}

The message of \prop{p:worst} is that the uniform mixture $\rho_{G/H}$ reveals
the least information about $H$ among all mixtures of coset states $\ket{Ha}$.
The distribution on irreps $V$ described in \prop{p:char}, together  with the
uniform state on $V^H$, also reveals the least information about $H$ among all
such distributions.

\section{A general algorithm}

In this section we will discuss a general algorithm for coherent HSP for an
arbitrary finite group $G$ and an arbitrary subgroup $H$.  It is an interesting
abstract presentation of all of the algorithms for dihedral groups in this
paper.  Unfortunately it might not be directly useful for any groups other than
dihedral groups.  

The algorithm uses the definitions and methods of \sec{s:character}, together
with a generalized notion of summand extraction.   In general if $V$
and $W$ are two unitary representations of $G$, their tensor product
decomposes as an orthogonal direct sum of irreps with respect
to the diagonal action of $G$:
\eq{e:extx}{V \tensor W \cong \bigoplus_X \cH_X^{W,V} \tensor X.}
Here again the direct sum runs over one representative of each isomorphism
class of irreps.  The Hilbert space $\cH_X^{W,V}$ is the \emph{multiplicity
factor} of the decomposition; its dimension is the number of times that $X$
arises as a summand of $V \tensor W$. The decomposition defines a partial
measurement of the joint Hilbert space $V \tensor W$, which extracts $X$ (and
$\cH_X^{W,V}$).  If $V$ and $W$ carry purely $H$-invariant states, then the
state on $X$ is also purely $H$-invariant.

\begin{algorithm} Input: An oracle that produces $\rho_{G/H}$.
\begin{description}
\item[1.] Make a list $L$ of copies of $\rho_{G/H}$.
Extract an irrep $V$ with a purely $H$-invariant state
from each copy.
\item[2.] Choose an objective function $\alpha$ on $\Irrep(G)$, the set of
irreps of $G$.
\item[3.] Find a pair of irreps $V$ and $W$ in $L$ such that $\alpha(V)$ and
$\alpha(W)$ are both low, but such that $\alpha$ is significantly higher for at
least one summand of $V \tensor W$.  Extract an irreducible summand $X$ from $V
\tensor W$ and replace $V$ and $W$ in $L$ with $X$.  Discard the
multiplicity factor.
\item[4.] Repeat step 3 until $\alpha$ is maximized on some irrep $V$.
Perform tomography on $V$ to reveal useful information about $H$.
\item[5.] Repeat steps 2-4 to fully identify $H$.
\end{description}
\label{a:gen} \end{algorithm}

For any given group $G$, \alg{a:gen} requires subalgorithms to compute the
character measurement \eqref{e:burnside} and the tensor decomposition
measurement \eqref{e:extx}.  Efficient algorithms for character measurements
and character transforms are a topic of active research
\cite{Beals:fourier,MRR:fourier} and are unknown for many groups.  We observe
that tensor decomposition measurement at least reduces to the character
measurement:

\begin{proposition} Let $V$ and $W$ be irreducible representations of a finite
group $G$.  If group operations in $G$ and summand extraction from $\C[G]$ are
both efficient, then summand extraction from $V \tensor W$ is also efficient.
\label{p:tensor}
\end{proposition}

\begin{proof} Embed $V$ and $W$ into separate copies of $\C[G]$ in a
$G$-equivariant way.  Then apply the unitary operator
$$U(\ket{a} \tensor \ket{b}) = \ket{b^{-1}a} \tensor \ket{b}$$
to $\C[G] \tensor \C[G]$.  The operator $U$ transports left multiplication by
the diagonal subgroup $G_\Delta \subset G \times G$ to left multiplication by
$G$ on the right factor.  Then summand extraction from the right factor of
$\C[G] \tensor \C[G]$ is equivalent to summand extraction from $V \tensor W$,
since, after $U$ is applied, the group action on the right factor of $\C[G]
\tensor \C[G]$ coincides with the diagonal action on $V \tensor W$.
\end{proof}

In light of Beals' algorithm to compute a character transform on the symmetric
group \cite{Beals:fourier} and \prop{p:tensor}, \alg{a:gen} may look promising
when $G = S_n$ is the symmetric group.  But the algorithm seems to work poorly
for this group, because the typical irrep $V$ of $S_n$ is very large. 
Consequently the decomposition \eqref{e:extx} typically involves many irreps of
$S_n$.  This offers very little control for a sieve.

Note that if \alg{a:gen} were useful for the symmetric group, its time
complexity would be $2^{O(\sqrt{\log |G|})}$ at best.  This is the same
complexity class as a known classical algorithm for the graph isomorphism or
automorphism problem \cite{BL:colored}, which is the original motivation for
the symmetric hidden subgroup problem (SHSP).  We believe that general SHSP is
actually much harder than graph isomorphism. If graph isomorphism does admit a
special quantum algorithm, it could be analogous to  a quantum polynomial time
algorithm found by Van Dam, Hallgren, and Ip \cite{vDHI:shift} for certain
special abelian hidden shift problems.  (In particular their algorithm applies
to the Legendre symbol with a hidden shift.) All of these problems have special
oracles $f$ that allow faster algorithms.

One reason that SHSP looks hard is that symmetric groups have many different
kinds of large subgroups.  For example, if $p_1, p_2, \ldots, p_n$ is a set of
distinct primes, then
$$D_{p_1p_2\ldots p_n} \injects S_{p_1 + p_2 + \cdots + p_n}$$
(exercise).  Thus DHSP reduces to SHSP.  Hidden shift in the symmetric group
also reduces to SHSP (exercise).

The sieve of \alg{a:gen} looks the most promising when the group $G$ is large
but $V \tensor W$ always has few terms. This is similar to demanding that most
or all irreps of $G$ are low-dimensional.  So suppose that all irreps have
dimension at most $k$ and consider the limit $|G| \to \infty$  for fixed $k$.
Passman and Isaacs \cite{IP:bounded} showed that there is a function $f(k)$
such that if all irreps have dimension at most $k$, then $G$ has an abelian
subgroup $\exp(A)$ of index at most $f(k)$.  By the reasoning of
\prop{p:arbhide}, the hardest hidden subgroup $H$ for a such a $G$ is one which
is disjoint from $\exp(A)$ (except for the identity).  But by the reasoning of
\sec{s:shifts}, any such hidden subgroup problem reduces to the hidden shift
problem on $A$.  The generalized sieve of \alg{a:gen} is not as fast as the
dihedral sieve on $D_A$.

\bibliography{qp,shared}

\providecommand{\bysame}{\leavevmode\hbox to3em{\hrulefill}\thinspace}
\begin{thebibliography}{10}

\bibitem{AKS:lattice}
Mikl\'os Ajtai, Ravi Kumar, and Dandapani Sivakumar, \emph{A sieve algorithm
  for the shortest lattice vector problem}, Proceedings of the thirty-third
  annual ACM symposium on Theory of computing, 2001, pp.~601--610.

\bibitem{Artin:algebra}
Michael Artin, \emph{Algebra}, Prentice Hall Inc., 1991.

\bibitem{BL:colored}
L\'aszl\'o Babai and Eugene~M. Luks, \emph{Canonical labeling of graphs},
  Proceedings of the fifteenth annual {ACM} symposium on theory of computing,
  ACM Press, 1983, pp.~171--183.

\bibitem{Beals:fourier}
Robert Beals, \emph{Quantum computation of {Fourier} transforms over symmetric
  groups}, {ACM} Symposium on Theory of Computing, 1997, pp.~48--53.

\bibitem{BKW:learning}
Avrim Blum, Adam Kalai, and Hal Wasserman, \emph{Noise-tolerant learning, the
  parity problem, and the statistical query model}, J. ACM \textbf{50} (2003),
  no.~4, 506--519, \mbox{arXiv:cs.LG/0010022}.

\bibitem{BDHHMSdW:distinctness}
Harry Buhrman, Christoph D\"urr, Mark Heiligman, Peter H{\o}yer, Fr\'ed\'eric
  Magniez, Miklos Santha, and Ronald de~Wolf, \emph{Quantum algorithms for
  element distinctness}, {IEEE} Conference on Computational Complexity, 2001,
  \mbox{arXiv:quant-ph/0007016}, pp.~131--137.

\bibitem{EH:noncomm}
Mark Ettinger and Peter H{\o}yer, \emph{On quantum algorithms for
  noncommutative hidden subgroups}, Adv. in Appl. Math. \textbf{25} (2000),
  no.~3, 239--251, \mbox{arXiv:quant-ph/9807029}.

\bibitem{EHK:almost}
Mark Ettinger, Peter H{\o}yer, and Emanuel Knill, \emph{Hidden subgroup states
  are almost orthogonal}, \mbox{arXiv:quant-ph/9901034}.

\bibitem{GSVV:quantum}
Michelangelo Grigni, Leonard~J. Schulman, Monica Vazirani, and Umesh~V.
  Vazirani, \emph{Quantum mechanical algorithms for the nonabelian hidden
  subgroup problem}, {ACM} Symposium on Theory of Computing, 2001, pp.~68--74.

\bibitem{HRT:normal}
Sean Hallgren, Alexander Russell, and Amnon Ta-Shma, \emph{Normal subgroup
  reconstruction and quantum computation using group representations}, {ACM}
  Symposium on Theory of Computing, 2000, pp.~627--635.

\bibitem{Hoyer:personal}
Peter H{\o}yer, 2003, personal communication.

\bibitem{IP:bounded}
I.~M. Isaacs and D.~S. Passman, \emph{Groups with representations of bounded
  degree}, Canad. J. Math. \textbf{16} (1964), 299--309.

\bibitem{IMS:instances}
G\'abor Ivanyos, Fr\'ed\'eric Magniez, and Miklos Santha, \emph{Efficient
  quantum algorithms for some instances of the non-abelian hidden subgroup
  problem}, \mbox{arXiv:quant-ph/0102014}.

\bibitem{Kitaev:abelian}
Alexei Kitaev, \emph{Quantum measurements and the abelian stabilizer problem},
  \mbox{arXiv:quant-ph/9511026}.

\bibitem{Kuperberg:dhspsim.py}
Greg Kuperberg, \texttt{dhspsim.py}, included with the source of
  arXiv:quant-ph/0302112.

\bibitem{MRR:fourier}
Cristopher Moore, Daniel Rockmore, and Alexander Russell, \emph{Generic quantum
  fourier transforms}, Proceedings of the fifteenth annual {ACM-SIAM} symposium
  on Discrete algorithms, SIAM, 2004, \mbox{arXiv:quant-ph/0304064},
  pp.~778--787.

\bibitem{MZ:exact}
Michele Mosca and Christof Zalka, \emph{Exact quantum fourier transforms and
  discrete logarithm algorithms}, Int. J. Qauntum Inf. \textbf{2} (2004),
  no.~1, 91--100, \mbox{arXiv:quant-ph/0301093}.

\bibitem{NC:book}
Michael~A. Nielsen and Isaac~L. Chuang, \emph{Quantum computation and quantum
  information}, Cambridge University Press, Cambridge, 2000.

\bibitem{Regev:dhsp}
Oded Regev, \emph{A subexponential time algorithm for the dihedral hidden
  subgroup problem with polynomial space}, \mbox{arXiv:quant-ph/0406151}.

\bibitem{Regev:quantum}
\bysame, \emph{Quantum computation and lattice problems}, SIAM J. Comput.
  \textbf{33} (2004), no.~3, 738--760, \mbox{arXiv:cs.DS/0304005}.

\bibitem{Serre:finite}
Jean-Pierre Serre, \emph{Linear representations of finite groups}, Graduate
  Texts in Mathematics, vol.~42, Spring-Verlag, 1977.

\bibitem{Shor:factorization}
Peter~W. Shor, \emph{Polynomial-time algorithms for prime factorization and
  discrete logarithms on a quantum computer}, SIAM J. Comput. \textbf{26}
  (1997), no.~5, 1484--1509, \mbox{arXiv:quant-ph/9508027}.

\bibitem{Simon:power}
Daniel~R. Simon, \emph{On the power of quantum computation}, SIAM J. Comput.
  \textbf{26} (1997), no.~5, 1474--1483.

\bibitem{vDHI:shift}
Wim van Dam, Sean Hallgren, and Lawrence Ip, \emph{Quantum algorithms for some
  hidden shift problems}, 2003, \mbox{arXiv:quant-ph/0211140}, pp.~489--498.

\end{thebibliography}

\end{document}